\documentclass[12pt]{article}
\usepackage{graphicx}

\newcommand{\bea}{\begin{eqnarray}}
\newcommand{\eea}{\end{eqnarray}}
\newcommand{\simgt}{\hbox{ \raise3pt\hbox to 0pt{$>$}\raise-3pt\hbox{$\sim$} }}
\newcommand{\simlt}{\hbox{ \raise3pt\hbox to 0pt{$<$}\raise-3pt\hbox{$\sim$} }}

\begin{document}
\begin{titlepage}
\title{\vspace{30mm}
Toward Determination of $m_t$\\
at 50~MeV Accuracy\thanks{Talk given 
at the Linear Colliders Workshop 2000, 
Fermilab, USA, October 2000.}
\vspace{20mm}
}
\author{
Y.~Sumino
\\ \\ 
    Department of Physics, Tohoku University\\
    Sendai, 980-8578 Japan
}
\date{}
\maketitle
\thispagestyle{empty}
\vspace{-4.4truein}
\begin{flushright}
{\bf TU--609}\\
{\bf January 2001}
\end{flushright}
\vspace{3.0truein}
\vspace{4cm}
\begin{abstract}
\noindent
{\small
The top quark mass will be determined to high accuracy from
the shape of the $t\bar{t}$ total production cross section 
in the threshold region at a future linear $e^+e^-$ collider.
Presently the estimated statistical error in the measurement of
$m_t^{\overline{\rm MS}}(m_t^{\overline{\rm MS}})$
is $\sim$50~MeV, while the estimated theoretical error is
150--200~MeV.
In order to reduce the theoretical
uncertainty to below 50~MeV,
we have recently
computed an important part of the higher-order corrections.
We deomonstrate the significance of the calculated correction.
}
\end{abstract}
\vfil

\end{titlepage}
  
\section{Introduction}

In the last Linear Collider Workshop held in Sitges (LCWS '99),
significant theoretical developments were reported concerning
the top quark physics in the threshold region: 
the next-to-next-to-leading
order corrections to the $e^+e^- \to t\bar{t}$ threshold cross
section have been computed \cite{NNLO1,NNLO2},
and understanding of the renormalon cancellation mechanism in
this cross section led to an improvement of convergence of
the perturbative expansion \cite{renormalon}.
These developments 
enabled a precise determination of the $\overline{\rm MS}$
mass of the top quark,
$\overline{m}_t \equiv m_t^{\overline{\rm MS}}(m_t^{\overline{\rm MS}})$,
in the future experiment at an $e^+e^-$ linear collider in the $t\bar{t}$
threshold region.

The top quark mass will be determined to high accuracy from
the shape of the $t\bar{t}$ total production cross section 
in the threshold region \cite{exp}.
The location of a sharp rise of the cross section is determined
mainly from the mass of the lowest-lying ($1S$) $t\bar{t}$ resonance.
Since the mass of the resonance can be
calculated as a function of the top quark mass and $\alpha_s$
from perturbative QCD,
we will be able to extract
the top quark mass from the measurement of the $1S$ peak position
of the cross section.

In the last workshop,
uncertainties in the determination of the top quark
mass were also estimated.
The statistical error in the measurement
of the $1S$ peak position of the $t\bar{t}$ threshold cross section
was estimated to be $\sim$50~MeV corresponding to 
a moderate integrated luminosity of 30~fb$^{-1}$ \cite{exp2}, while
the theoretical uncertainty in the relation between the
peak position and the top quark mass was estimated to be
150--200~MeV \cite{topcollab}.

The motivation of the present study is to reduce the theoretical
uncertainty to the level of the 
experimental error (i.e.\ $\simlt 50$~MeV).
We have computed an important piece of the higher-order corrections
in order to achieve this goal.
More precisely,
Kiyo and the present author 
have computed in \cite{ks} the ${\cal O}(\alpha_s^5 m)$
correction to the quarkonium $1S$ spectrum in the large-$\beta_0$
approximation.\footnote{
The same result was obtained later in \cite{mceff}.
}
We clarify the significance of the correction in the context of
the measurement of the top quark mass.

\section{Renormalon Cancellation in the Quarkonium Spectrum}

\begin{figure}[tbp]
  \hspace*{\fill}
    \includegraphics[width=17cm]{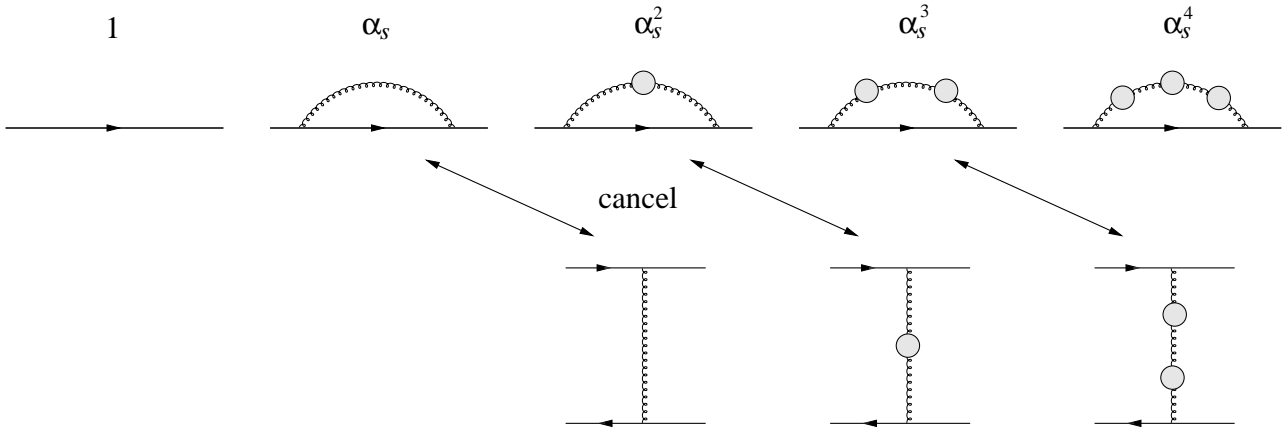}
  \hspace*{\fill}
  \\
  \hspace*{\fill}
\caption{\footnotesize
The figure showing how cancellations should take place between the
diagrams.
The orders of the potential graphs are shifted by one power of
$\alpha_S$ which is provided by the inverse of the Bohr radius
$\left< \frac{1}{r} \right> \sim \alpha_S m$.
      \label{shiftorder}
}
  \hspace*{\fill}
\end{figure}

The theoretical prediction of the quarkonium mass
spectrum is given as a 
series expansion in $\alpha_s$.
There seemed to be some confusions among theoreticians
in identifying the order of accuracy
of the theoretical prediction.
These originated from the  fact that
the renormalon cancellation
in the perturbation series of the quarkonium spectrum is realized
in a slightly non-trivial way.
When the pole mass and the binding energy are given as 
series expansions 
in $\alpha_S$, renormalon cancellation takes place between
the terms whose orders in $\alpha_S$ differ by one \cite{hlm}:
\bea
\begin{minipage}[c]{15cm}
\vspace{2mm}
  \begin{flushleft}
\begin{picture}(100,60)(0,0)
\put(60,42){$2 m_{\rm pole} = 2 \overline{m} \, \,
( 1 + A_1 \, \alpha_S + A_2 \, \alpha_S^2 + A_3 \, \alpha_S^3 
+ A_4 \, \alpha_S^4 + \cdots ),$}
\put(60,0){$E_{\rm bin} ~~~ = 2 \overline{m} \, \,
( ~~~~~~~~~~~~~~~~
B_2 \, \alpha_S^2 + B_3 \, \alpha_S^3 
+ B_4 \, \alpha_S^4 + \cdots ).$}
\put(235,23){\footnotesize cancel}
\put(190,24){\vector(-3,2){20}}
\put(190,24){\vector(3,-2){20}}
\put(275,24){\vector(-3,2){20}}
\put(275,24){\vector(3,-2){20}}
\put(230,24){\vector(-3,2){20}}
\put(230,24){\vector(3,-2){20}}
\put(320,24){\vector(-3,2){20}}
\put(320,24){\vector(3,-2){20}}
\end{picture}
\end{flushleft}
\vspace{2mm}
\end{minipage}
\label{cancel}
\eea
Intuitively this may be seen by comparing
the diagrams shown in Fig.~\ref{shiftorder}.
One way to understand the shift in the order counting is to
regard that
an extra power of $\alpha_S$ in the binding energy is provided
by the inverse of the Bohr radius 
$\left< r^{-1} \right> \sim \alpha_S m$,
\bea
\mbox{i.e.}~~~~~~~~~~~~~~
\left< C_F\frac{\alpha_S^n}{r} \right> \sim \alpha_S^{n+1} m .
\eea
Another way to understand this is as follows.
When the binding energy is expanded in $\alpha_s$, 
\bea
E_{\rm bin} = 2 \overline{m} \, \sum_{n=2}^{\infty}
P_n (\log \alpha_s ) \, \alpha_s^n ,
\eea
the coefficients $P_n(\log \alpha_s)$
are polynomials of $\log \alpha_s$, which is a
characteristic feature of the perturbative expansion of the boundstate
spectrum.
For large $n$, the polynomials behave as 
\bea
P_n(\log \alpha_s) \sim \alpha_s^{-1}, 
\eea
which effectively decrease
the power of $\alpha_s$ by one.

It is legitimate to consider that the present perturbative calculation of
the quarkonium spectrum, when expressed in terms of the quark
$\overline{\rm MS}$-mass, has a 
{\it genuine accuracy} at ${\cal O}(\alpha_S^3 m)$.
In fact, in formal power countings, the last known term in
the relation between the
$\overline{\rm MS}$-mass and the pole-mass of a quark is 
${\cal O}(\alpha_S^3 m)$ \cite{polemass2}, while the last known term of
the binding energy (measured
from twice of the quark pole-mass) is
${\cal O}(\alpha_S^4 m)$.
The former term
includes in addition to a genuine ${\cal O}(\alpha_S^3 m)$
part the leading renormalon contribution which does not become
smaller than ${\cal O}(\Lambda_{\rm QCD})$ \cite{bbb}.
This renormalon contribution is cancelled \cite{renormalon}
against the renormalon
contribution \cite{al} contained in the latter term.
Therefore, after cancellation of the leading renormalons,
the {\it genuine} ${\cal O}(\alpha_S^3 m)$ part of the mass relation
determines the accuracy of the present perturbation series
relating the quark $\overline{\rm MS}$-mass and the quarkonium
spectrum.

As stated, for the binding energy the calculation including the
{\it genuine} ${\cal O}(\alpha_S^4 m)$ corrections has already 
been completed.
Also, the 
``large-$\beta_0$ approximation''
\cite{bb}
is known to be a pragmatically feasible and empirically successful
estimation method of the leading renormalon contributions.
Taking these into account, one finds that it is sufficient
to calculate further the following two corrections
in order to improve the accuracy of
the spectrum by one order and to
achieve a genuine accuracy at ${\cal O}(\alpha_S^4 m)$:
(I) the ${\cal O}(\alpha_S^4 m)$ relation between the
$\overline{\rm MS}$-mass and the pole-mass, and
(II) the binding energy at ${\cal O}(\alpha_S^5 m)$ in the
large-$\beta_0$ approximation.
This is because the leading renormalon contribution in the full
${\cal O}(\alpha_S^5 m)$ correction to the binding energy will be
incorporated by the large-$\beta_0$ approximation and the
remaining part is expected to be irrelevant at ${\cal O}(\alpha_S^4 m)$.

Of these two corrections we have calculated
(II) analytically for the $1S$-state in \cite{ks}.
In the next section we will examine the series expansion of
the ``toponium'' $1S$ spectrum.
Also we will check validity of the above general argument
explicitly at ${\cal O}(\alpha_S^3 m)$ where we know the
exact result.

\section{``Toponium'' $1S$ State Spectrum}

Taking the input parameter as $m_{t,{\rm pole}} = 174.79$~GeV/%
$\overline{m}_t = 165.00$~GeV and setting $\mu = \overline{m}_t$
[i.e.\ expansion parameter is $\alpha_S(\overline{m}_t)=0.1092$],
we obtain the series expansions of the mass spectrum of
the toponium $1S$ state:
\bea
M_{1S} &=& 
2 \times ( 174.79 - 0.46 - 0.39 - 0.28 - 0.19^* )~{\rm GeV}
~~~~~
\mbox{(Pole-mass scheme)}
\\
&=&
2 \times ( 165.00 + 7.21 + 1.24 + 0.22 + 0.052^* )~{\rm GeV}
~~~~~
\mbox{($\overline{\rm MS}$-scheme)} .
\label{mtmsbar}
\eea
The numbers with stars ($*$) are calculated in the
large-$\beta_0$ approximation (in the case of $\overline{\rm MS}$-scheme
the ${\cal O}(\alpha_s^4 m)$ term of the
pole-$\overline{\rm MS}$ mass relation
is replaced by the large-$\beta_0$ approximation as well).
The series expansion converges very slowly when the $1S$ spectrum
is expressed using the pole mass.
The series converges much faster when the spectrum is 
expressed by the $\overline{\rm MS}$ mass.

As we argued in the previous section, parametric accuracy of the last
term in (\ref{mtmsbar}) 
is ${\cal O}(\alpha_S^4 m)$ and we need to know further only
the exact 
${\cal O}(\alpha_s^4 m)$ term of the pole-$\overline{\rm MS}$ mass relation
to make a perturbative evaluation accurate
up to this order (the exact form of 
the binding energy at ${\cal O}(\alpha_s^5 m)$ is not necessary).
In order to verify validity of this argument, we replace 
the binding energy at ${\cal O}(\alpha_s^4 m)$ by its
value in the large-$\beta_0$ approximation.
Then the ${\cal O}(\alpha_S^3 m)$ term of 
(\ref{mtmsbar})
changes to 0.23.
Thus, we do not lose accuracy at this order by the replacement.
On the other hand, if we replace in addition
the ${\cal O}(\alpha_s^3 m)$ term of the
pole-$\overline{\rm MS}$ mass relation by its
values in the large-$\beta_0$ approximation, the same
term changes to 0.31.
Thus, we lose the accuracy at ${\cal O}(\alpha_S^3 m)$.
These aspects are consistent with our general argument.
Also, they suggest that the last term
of (\ref{mtmsbar})
would be a reasonable estimate of the order of magnitude of
the exact ${\cal O}(\alpha_S^4 m)$ term.
Then, when the 
${\cal O}(\alpha_s^4 m)$ term of the pole-$\overline{\rm MS}$ mass relation
is calculated, the remaining theoretical uncertainty is expected
to be below 50~MeV.

\begin{figure}[tbp]
  \hspace*{\fill}
    \includegraphics[width=9cm]{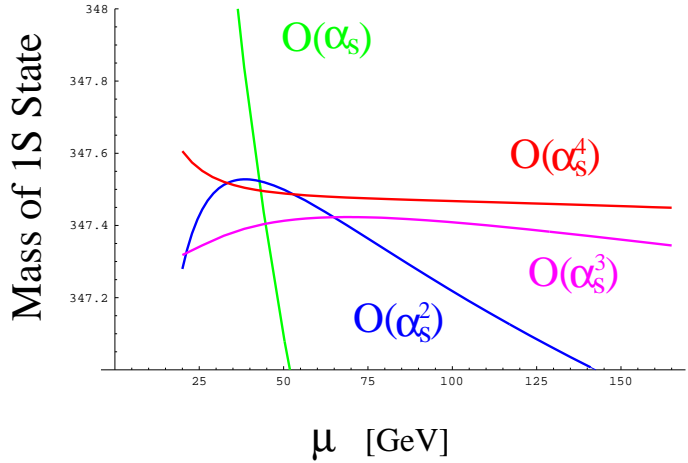}
  \hspace*{\fill}
  \\
  \hspace*{\fill}
\caption{\footnotesize
$\mu$-dependences of the sums of the series expansion for
the $1S$ spectrum.
The ${\cal O}(\alpha_s^4)$ term is evaluated in the large-$\beta_0$
approximation.
      \label{mu-dep}
}
  \hspace*{\fill}
\end{figure}

Finally let us examine the dependence of the $1S$ spectrum
$M_{1S}(\alpha_s(\mu),\overline{m}_t)$ on the scale $\mu$.
As shown in Fig.~\ref{mu-dep} the scale dependence decreases as we include
more terms of the series expansion.
The scale dependence of the sum of the series up to
${\cal O}(\alpha_s^4 m)$, where the last term is evaluated in
the large-$\beta_0$ approximation, is also consistent with
the estimate of the theoretical uncertainty below 50~MeV.

\section{Discussions}

There are corrections, other than the
4-loop pole-$\overline{\rm MS}$ mass relation, 
that should be computed before we achieve 50~MeV
accuracy ultimately.
These are:
\begin{enumerate}
\item
The final-state interaction corrections (non-factorizable corrections)
to the $e^+e^- \to t\bar{t}$ total cross section.
\item
The electroweak corrections to the cross section.
These include the contributions from the tree-level non-resonant-type
diagrams and the one-loop resonant-type corrections.
\end{enumerate}
In addition, the uncertainty originating from the large
next-to-next-to-leading order corrections to the normalization
of the cross section affects the measurement of the top quark mass.
This problem should be solved as well.
An attempt has been given recently in \cite{rg}.

\section*{Acknowledgements}

This work is based on the collaborations with Y.~Kiyo and
T.~Nagano.
The author was supported in part
by the Japan-German Cooperative Science
Promotion Program.

\end{document}